# Technical Report on Deploying a highly secured OpenStack Cloud Infrastructure using BradStack as a Case Study. *

[1]Bashir Mohammed, [1]Sibusiso Moyo, [1]K. M Maiyama, [1]Sulayman Kinteh, [1]Al Noaman M.K. Al-Shaidy, [1]M. A. Kamala and [2]M. Kiran

[1]*Cloud Computing Modelling and Simulation Research Group*
*School of Electrical Engineering and Computer Science*
*University of Bradford.UK*
b.mohammed1@bradford.ac.uk
s.moyo10@ bradford.ac.uk
k.m.maiyama@ bradford.ac.uk
s.kinteh@bradford.ac.uk
a.n.m.alshaidy@bradford.ac.uk
m.a.kamala@bradford.ac.uk

[2]*Lawrence Berkeley National Labs, California, USA*
mkiran@es.net

30th of October 2017

**Abstract**

Cloud computing has emerged as a popular paradigm and an attractive model for providing a reliable distributed computing model.it is increasing attracting huge attention both in academic research and industrial initiatives. Cloud deployments are paramount for institution and organizations of all scales. The availability of a flexible, free open source cloud platform designed with no propriety software and the ability of its integration with legacy systems and third-party applications is fundamental. Open stack is a free and opensource software released under the terms of Apache license with a fragmented and distributed architecture making it highly flexible. This project was initiated and aimed at designing a secured cloud infrastructure called BradStack, which is built on OpenStack in the computing Laboratory at the University of Bradford. In this report, we present and discuss the steps required in deploying a secured BradStack Multi-node cloud infrastructure and conducting Penetration testing on OpenStack Services to validate the effectiveness of the security controls on the BradStack platform. This report serves as a practical guideline, focusing on security and practical infrastructure related issues. It also serves as a reference for institutions looking at the possibilities of implementing a secured cloud solution.

---

# Table of Contents

# 1  Introduction

Cloud Computing represents a major shift in the Information Technology (IT) services landscape. Cloud has simplified the availability of enterprise-grade computing power to the organisation, without the need to invest in hardware or staff and associated costs for procurement activities. According to Mell and Grance [1], Cloud refers to the use of computing resources in which hardware and/or software resides on a remote machine often based on virtualization, computing technologies distributed and delivering to the end users as a service over a network. The most popular delivery network is being the internet, for example, when using online services to edit documents, listen to music, send emails and store files (e.g. Dropbox). Cloud architectures are highly abstract resources, scalable, flexible, near instantaneous provision, shared resources on demand, usually with a 'pay as you go' billing system, which turns computing into a utility.

The adoption of cloud computing by organisations has become prevalent in this 21st century. A recent survey conducted by North Bridge shows that 50% of the companies had implemented cloud about 90% of which, their operations in one way or another is compared to only 60% in 2014 [2].Furthermore, Gartner[3] projected that cloud market will reach $246.8 billion in 2017 alone, and also forecasted it to grow by more than 34.7% to reach $332.7 billion in 2019, where 55.3% expected to reach $383.3 billion by 2020 respectively [3]. Cloud computing provides numerous advantages to companies; amongst them are low cost, flexibility, faster market time, and faster deployment. However, security of data is often cited as one of the major concerns for those considering a move towards cloud-based services [4].

Failure to provide an appropriate security control or mechanism on one's infrastructure could lead to data breaches, which could result in financial losses, loss of reputation and reduced customer confidence. Thus, cloud security is one of the main problems in cloud computing [5]. As the above projection shown by Garter[3], more organizations will be using cloud computing, which emphasizes the need for more cloud security research to be conducted to find innovative ways to secure cloud infrastructure.

The purpose of this project is to re-deploy and secure the BradStack environment whilst addressing the known vulnerabilities from previous architectural design including the network infrastructure to ensure that the setup is secure. BradStack is a customized implementation of OpenStack open source Cloud Computing Platform (CCP) that focuses on the deployment of main services such as Infrastructure as a Service (IaaS), Platform as a Service (PaaS) and Software as a Service (SaaS). Cloud Modelling and Simulation research group (CMSRG) at University of Bradford developed BradStack Cloud Computing Platform (CCP) project. The project mentioned above aimed at providing cloud resources mainly used for research as well as other services needed by the University community and beyond.

## 1.0 The Concept of Cloud and it's Deployment Models

Cloud computing can be referred to as a model for enabling appropriate, ubiquitous, on-demand network access to a public pool of configurable computing assets (e.g., networks, storage, servers, services and applications). These assets can be quickly provisioned and released with negligible management effort or service provider communication [1]. Cloud computing can be categorized into three distinct elements. Each element has a purpose which performs specific tasks as follows; data centers, distributed servers, and clients[6].The four main cloud deployment models are public, private, community and hybrid.

- Public clouds are provided to the public or a large industry group. This is managed by a third party selling cloud services. As cloud technology develops, public cloud services are becoming more attractive to Business companies as well. Many actors such as critical infrastructure providers, including financial institutions has shown special interest in cloud[7].

- Private cloud is owned and managed by a single organization that concentrates on controlling the mechanism of visualizing resources and automating services that are used and customized by various lines of business and constituent groups. A private cloud provides services to an organization through an intranet. Private clouds can be linked to each other to form a partner cloud. Private clouds are operated solely for an organization[5].

- Community cloud it is dedicated or allows services to be available to a professional community or group of organization that comprises of Subcontractors, Branches, allies and so forth to operate collaboratively on a project. It may also be a government cloud dedicated to state establishments [8].

- Hybrid cloud model is a mixture or combination of private and public cloud infrastructures working together. As a result, hybrid cloud inherits the properties of both private cloud and public cloud. It allows organizations to manage their critical data and applications in private while outsourcing other non-critical activities to the public cloud[9][10].

## 1.1 Cloud Service Models

There are three levels of essential services offered by cloud computing: Infrastructure as a service (IaaS), platform as a service (PaaS) and software as a service (SaaS).

- Infrastructure as a service (IaaS), is the most basic and important cloud service model under which virtual machines, load balancers, fault tolerance, firewalls and networking services are provided[11]. The client or cloud user, is provided with capability to provision processing, storage, networks and other fundamental computing resources, to deploy and run arbitrary software such as operating system and applications. Common examples of these services include Rackspace, GoGrid, EC2, Google Apps, Concur, Cisco Webex, Citrix GoTo Meetings, Adobe Marketing Cloud, Facebook, Flickr) and Amazon cloud[12] [13] .

- Under the PaaS model, a computing platform including APIs, operating system and development environments are provided as well as programming language execution environment and web servers. The client maintains the applications, while the cloud provider maintains the service run times, databases, server software, integrated server oriented architectures and storage networks. Various types of PaaS vendors offerings can include complete application hosting, development, testing and extensive integrated services that include scalability and maintenance[14]. Some key players include Microsoft Windows Azure and Google Apps engine GoDaddy, Windows Azure, Apprenda, Google App Engine, Amazon Web Services, and WordPress. The main benefit of these services include focus on high value software rather than infrastructure, leverage economies of scale and provide scalable go-to-market capability [15].

- SaaS provides clients the capability to use provider application executing on a cloud infrastructure. An entire application is available remotely and accessible from multiple client devices through thin client interfaces such as web browsers. Cloud user do not manage or control the underlying cloud infrastructure [2] but providers install and operate the application software. Example providers for this service include Salesforce, Facebook and Google Apps, Amazon EC2, Rackspace, Microsoft Azure, Google Compute Engine and Amazon Web Services [15]–[17].

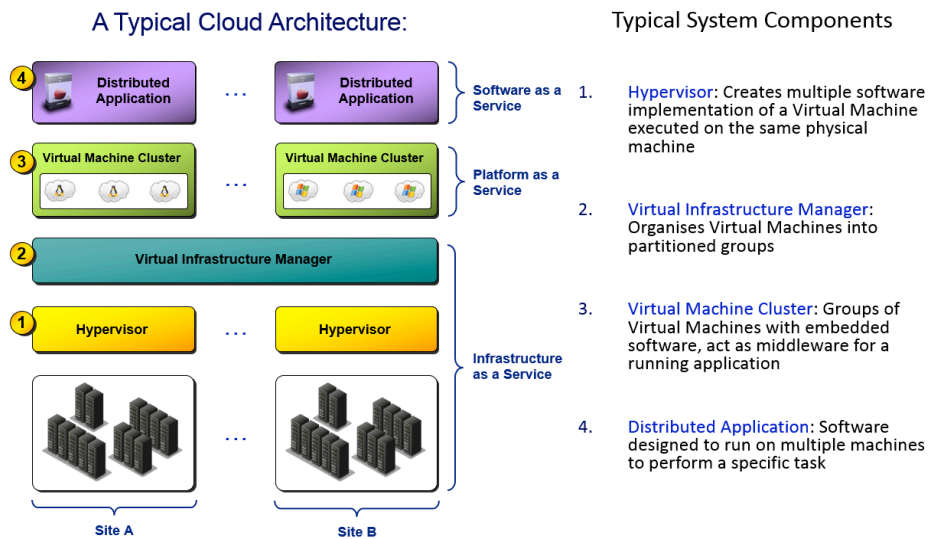

Figure 1.0 Cloud Architecture[45]

## 2 OpenStack Cloud Platform Overview

OpenStack, an open source cloud operating system is a cloud operating system that controls large pools of compute, storage and networking resources throughout a datacenter. The resources are managed through a dashboard giving administrators control while empowering its users to provision resources through web interfaces [18], [19].It is designed to provide flexibility to design private cloud, with no proprietary hardware or software requirements and the ability to integrate with legacy systems and third-party technologies. It is designed to manage and automate pools of compute resources to work with virtualization technologies and high-performance computing configurations. Administrators deploy OpenStack compute using one of the multiple supported hypervisors in a virtualized environment. KVM and XenServer are popular choices for hypervisor technology and recommended for most use cases. Linux container technology such as LXC is also supported for scenarios where users wish to minimize virtualization overhead to achieve greater efficiency and performance. In addition to different hypervisors, OpenStack supports ARM based processors and alternative hardware architectures [18]–[20].

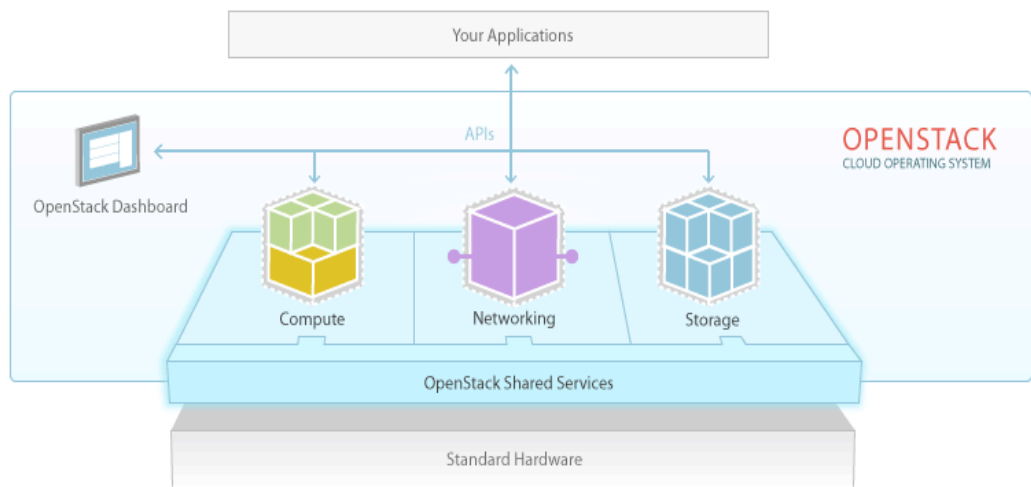

Figure 2.1 OpenStack overview[21]

The OpenStack software consists of the Compute, Network and Storage services which collectively provide the OpenStack cloud management capabilities as shown in Figure 2.1. These pools of resources can be managed or consumed from a single place using the OpenStack dashboard. Both the users and the administrators can utilize the dashboard to perform their tasks in a simple easy manner[22]. OpenStack cloud operating system controls the layer that sits above all the virtualized layers and provides a simple and consistent way to access services regardless of the technology used on the hypervisor. OpenStack Hypervisor technology list (KVM, Hyper V, VMware, Xen, etc.)

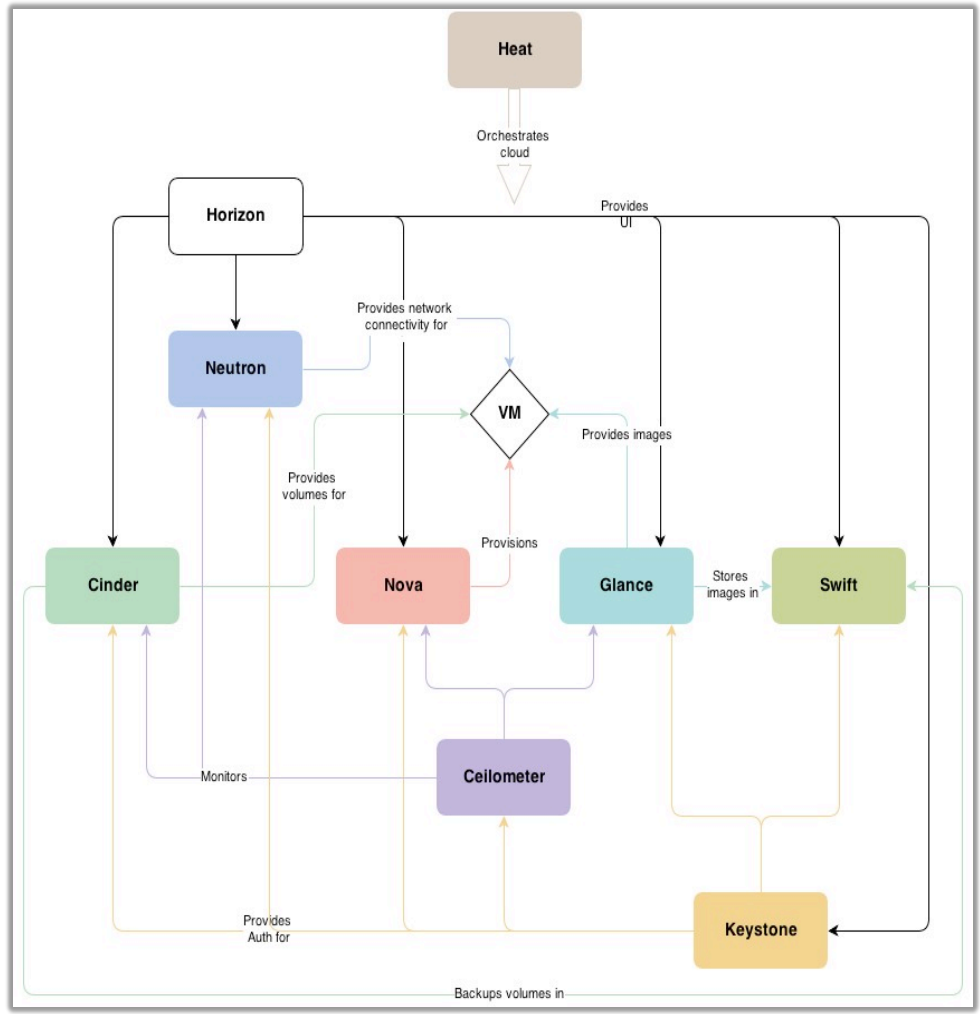

Figure 2.2 OpenStack Architecture[22]

Figure 2.2 presents the OpenStack conceptual architecture while Table 1 describes the OpenStack services that make up the OpenStack architecture: OpenStack consists of 6 main components, which apply to infrastructure as a service(IaaS), these are OpenStack Identity Service (Keystone), OpenStack Compute (Nova), OpenStack Networking service (neutron), OpenStack Image Service (Glance), OpenStack Block Storage service (cinder) and OpenStack Dashboard (Horizon), [22]. This work only focuses on Keystone, Nova, Neutron, Glance, Cinder and Horizon, due to the time and document constants.

Table 1  OpenStack Services.

| | Service | Project Name | Brief Description |
|---|---|---|---|
| **CLOUD MANAGEMENT** | Compute | Nova | This Project manages the lifecycle of compute instances in an OpenStack environment. Its responsibilities include spawning, scheduling and decommissioning of virtual machines on demand. |
| | Networking | Neutron | The Neutron Enables Network-Connectivity-as-a-Service for other OpenStack services, such as OpenStack Compute. Provides an API for users to define networks and the attachments into them. Has a pluggable architecture that supports many popular networking vendors and technologies. |
| | Dashboard | Horizon | Horizon Provides a web-based self-service portal to interact with underlying OpenStack services, such as launching an instance, assigning IP addresses and configuring access controls. |
| **SHARED SERVICES** | Identity service | Keystone | The keystone Provides an authentication and authorization service for other Open-Stack services. Provides a catalogue of endpoints for all OpenStack services. |
| | Image service | Glance | Glance simply Stores and retrieves virtual machine disk images. OpenStack Compute makes use of this during instance provisioning. |
| | Telemetry | Ceilometer | Ceilometer Monitors and meters the OpenStack cloud for billing, benchmarking, scalability, and statistical purposes. |
| **STORAGE** | Block Storage | Cinder | Provides persistent block storage to running instances. Its pluggable driver architecture facilitates the creation and management of block storage devices. It supports many popular networking vendors and technologies. |
| | Object Storage | Swift | Swift stores and retrieves arbitrary unstructured data objects via a RESTful, HTTP based API. It is highly fault tolerant with its data replication and scale out architecture. Its implementation is not like a file server with mountable directories. |
| **HIGH LEVEL SERVICES** | Database service | Trove | Trove provides scalable and reliable Cloud Database-as-a-Service functionality for both relational and non-relational database engines. |
| | Orchestration | Heat | The Heat Project Orchestrates multiple composite cloud applications by using either the native *HOT* template format or the AWS Cloud Formation template format, through both an OpenStack-native REST API and a Cloud Formation- compatible Query API. |

# 3  Open Source Cloud Platform Comparison

This section presents a critical analysis of open source toolkits and their suitability as well as the findings obtained from the comparative studies. The methodology and approach based on the concept of cloud computing is presented and the experimental results for the benefit of migrating to cloud using OpenStack.

## 3.1 Findings obtained from the Comparative Technical Studies

Table 2 presents a comparative technical analysis between various open source IaaS cloud computing solutions such as Eucalyptus, CloudStack, OpenStack, OpenNebula, Nimbus, Xen Cloud Platform (XCP), OpenIoT and AbiCloud. Each open source software provides the IaaS to deliver a standard virtualization environment, using the following classification criteria: Origin and Community support, architecture, relation with Amazon, cloud implementation, programming and scripting language, Hypervisor supported, Operating system support, databases, Image management, VM migration support, Fault tolerance and load balancing. The findings obtained from the comparative technical studies carried out between the eight OpenSource Cloud solutions are as follows:

- OpenNebula is ideal for users who wants to setup a cloud environment using a couple of machines because of its architecture as a classical cluster, which is an architecture with a front-end and a set of cluster nodes to run the VMs. This implies that at least one physical network is required to connect all cluster nodes with the front-end. It is compatible with Xen, KVM and VMware hypervisors and works with various operating systems such as CentOS, Debian, fedora, RHEL and Ubuntu. It supports Fault tolerance and Load balancing.

- Eucalyptus is tailored towards institutions or organizations that want to build their own private cloud but does not support virtual machine migration. It is compatible with AWS application programming interface.

- Nimbus mainly targets features of the scientific community such as support for proxy credentials, batch schedulers, and best-effort allocations. It is also compatible with AWS application programming interface and most Linux distributions. However, it does not support virtual machine migration, making it more suitable for users interested in internal techniques of the system.

- XCP is an open source enterprise server virtualization and cloud-computing platform. It concentrates more on the provision of tools to manage a collection of virtualized host. Even though it was developed under Public License and supports Fedora, Red Hat and CentOS, its greatest limitation is that, it does not offer a user interface to manage the cloud user access using only command line input.

- CloudStack supports key features such as Hypervisor agnostic, built high availability for host, VMs, snapshot management, Virtual machine migration, load balancing and fault tolerance. It is compatible with AWS's application

- programming interface and its programming framework in Java and Python. The biggest drawback is that CloudStack architecture is to have a reduced installation flexibility and lack of support for shared access.

- AbiCloud is an open source platform used to develop, manage and integrate private and public cloud in a homogenous environment. However, it does not support VM migration, fault tolerance and is not compatible with AWS's application programming interface even though its programming framework is Java and Python and works with Xen, KVM, VMware and VirtualBox.

- OpenStack is free and open source software released under the terms of the Apache license with a fragmented and distributed architecture making it highly flexible. It is compatible with hypervisors such as KVM, XEN, QEMU and Hyper-V and is written in Python and UNIX shell. It has AWS EC2 compatibility, supports AWS S3 API and is an open source platform for building private and public cloud. It has a broad range of support from major tech key industry players like Dell, IBM, NASA, Rackspace, Cisco, HP etc. OpenStack supports image management, VM migration, load balancing and fault tolerance as well as a web interface for user's access.

- OpenIoT is completely free open source blueprint middleware infrastructure for developing and integrating non-trivial IoT applications. It supports virtually any sensor type available by enabling and facilitating the integration and use of virtually any internet-connected physical or virtual sensor to IoT applications. It embraces AWS and it is an open source platform for building a sensor-centric private and public cloud. The OpenIoT Cloud software is written in Java and built on popular open source packages such as Apache Active MQ and JBoss Netty [23]. It is also available under a Business-Friendly License, i.e. GPL 3.0 which enables enterprises particularly SMEs to use it for solution development and deployment. It uses the web as an access interface.

Table 2 Comparison of Opensource Cloud implementation tools

| Cloud Computing Platform (developed by, year from) | License | Hypervisor supported | Architecture | Relation with Amazon | Cloud Implementation | Programming Framework | OS support | Database | Image mgt. (VM migration support) | Load balancing (Fault Tolerance) | Access interface |
|---|---|---|---|---|---|---|---|---|---|---|---|
| Eucalyptus (2008, University of California, Eucalyptus System Inc Cloud.com) | GPL (General public license) | XEN, KVM, VMWare | Cloud Controller, Walrus, Cluster Controller, Node Controller, Storage Controller | Embraces AWS-API | Open Source platform(OSP) for developing private clouds | Java, C, Python | GNU/Linux can host Linux and windows VMs | Postgre SQL | Managed by Euca2ools (No) | Cloud Controller (Cluster controller's separation) | EC2 WSAPI |
| CloudStack (2008, Cloud.com) | GPL (General public license) | Xen Server, Xen Cloud Platform, KVM, VMWare | Monolithic architecture | Embraces AWS-API | OSP for developing private and public clouds | Java, Python | GNU/Linux can host Linux and windows and BSD for guests VMs | MySQL | Managed by CloudStack (Yes) | Cloud Controller (Replication) | Rich management brandable self-service user interface |
| OpenStack (2010, Rackspace, NASA) | Apache License V2 | VMWare, ESX & ESXL, Microsoft hyperV, Xen KVM, Virtual Box | Compute (Nova) Object Storage (Swift) Image Service (Glance) | Embraces AWS-API AWS EC2 API AWS S3 API | OSP for developing private, public cloud | Python, XML, JavaScript | CentOS, Debian, Fedora, RHEL openSUSE, SLES and ubuntu | SQLite3, MySQL & Postgre | OpenStack image Service (Glance) (Yes) | Cloud Controller (Replication) | Web interface |
| OpenNebula (2005, European Union) | Apache License V2 | Xen, KVM, VMWare, vCenter | Classical cluster | Embraces AWS-API | OSP for developing private, public hybrid cloud | C++, C, Ruby, Java, Shell script | CentOS, Debian, Fedora, RHEL openSUSE, SLES and ubuntu | SQLite, MySQL | Data store or Image repository (Yes) | Nginx (Database backend (registers VM information)) | EC2 WS API OCCI API |
| Nimbus (2009, University of Chicago) | Apache License V2 | Xen +KVM | Virtual cluster | Embraces AWS-API | OSP for developing private, and community cloud | Java, Python | Most Linux distributions | Postgre SQL | Managed by Nimbus (No) | LeContext Broker (Periodic verification of Cloud nodes) | ECS WS API Nimbus WSRF |
| Xen Cloud Platform (XCP) (2012, Citrix Xen Server) | GPL (General public license) | Xen | Centralized three components minimum of 2 servers | Yes | Turnkey virtualization that provides out of the box visualization | OCaml language | Linus (Fedora, RedHat, CentOS, Wondows 7) | VastSky | Managed by Xen (Yes) | XAPI (VM states synchronization) | Command lines |
| AbiCloud (2009, Abiquo) | GPL (General public license) | Xen, Virtual Box, KVM, VMWare | Centralized three components minimum of 2 servers | No | OSP for developing private & public cloud | C++, Ruby, Java, Python | Linux (Ubuntu, CentOS, WindowsXP, MacOS) | HDFS | Managed by AbiCloud (No) | AbiServer (No) | Web interface & adobe Flex |
| OpenIoT (2013, European Union) | GPL (General public license) | Virtual Box, VMware Virtual Development Kit (VDK) | IoT Cloud Controller, Message Broker, Sensors, Clients. | Yes Embraces AWS-API, Sensor API, Client-API | OSP for developing Sensor-Centric private & public cloud | Java | Most Linux distributions Cross-Platform | MySQL | Managed by OpenIoT (No) | Cloud Controller (No) | Web interface |

## 3.2 Summary

Figure 3.1 presents the Open Source Cloud Tools Analysis Quadrant where it was observed that the tools located in third quadrant (OpenStack and OpenNebula) are the most flexible in terms of data centre virtualization, infrastructure provision and relation with Amazon web services while the least tools located in quadrant one (AbiCloud, Nimbus and Xen) and least are not as flexible as the others. Eucalyptus, OpenIOT and CloudStack are in the middle of the quadrant because they are not as flexible as tools located in the third quadrant but are much better that tools located in the first quadrant.

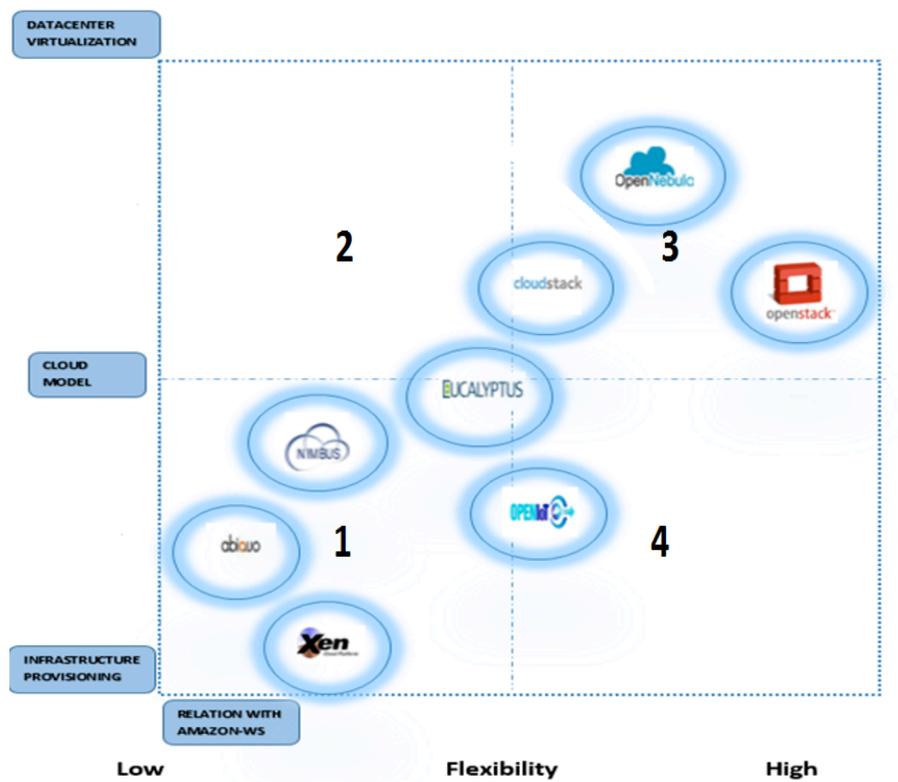

Figure 3.1 Open Source Cloud Tools Analysis Quadrant

# 4 Cloud information security vulnerabilities

The security risks and requirements for each cloud model vary depending on factors such as architecture, deployment model and the sensitivity of information assets. Security concerns from traditional infrastructure still affect the cloud infrastructure especially from the perspective of cloud service provider, including threats that are associated with cloud computing such as virtual machine escape [24].One of the first steps in reducing risk associated with cloud computing is to identify cloud security threats. Cloud Security Alliance (CSA) has created industry-wide standards for cloud security. Which provides best practices to secure cloud computing, that is addressed within the fourteen domains of CSA Guidance. CSA suggest a regular vulnerability scanning, prompt patch management, and can help mitigate such threats. While there are numerous security threats in the cloud computing related to on-demand, the shared nature of cloud computing. Data breaches, insecure API's and system application vulnerabilities will be explored in this research. The Open Web Application Security Project OWASP (2017) and CSA generated a list of the most critical security vulnerabilities; which are Buffer Overflow, Cross Site Scripting (XSS), Command Injection, SQL Injection, etc. Cloud computing heavily relies on the web application to deliver the service to its customers, therefore, is important to minimize the potential risk. Thus, this report will cover the most critical security flaws highlighted by OWASP and CSA. Furthermore, it has been addressed that insecure API in the cloud can affect the organization business, loss of customers, financial loss, and loss organization reputation [5].In 2016 a breach that exposed insecure API took place at an online greeting card vendor Moonping in USA. As a result, numerous customer's information was obtained. Hence, it obvious that APIs security is a serious security challenge because it is the public door to the cloud.

## 4.1 Overview of information security

Information systems security refers to any activities designed to protect information systems, protect sensitive information or data from unauthorized use, access, modification, destruction, disclosure and disruption in order provide to provide integrity, confidentiality and availability
The international standard, ISO/IEC 27002 [25], defines information security as the preservation of, confidentiality, integrity and availability.it consists of the processes and mechanisms that are deployed to protect information systems from all threats whether the external or internal. According to [26], [27]these terminologies are also known in information security as the CIA triangle.

### 4.1.1 Confidentiality

[28] defines confidentiality as the prevention of sensitive information from reaching unauthorized entity. End-users who used the cloud services to store their personal data in the provider's infrastructure, concern about user data confidentiality and privacy. Therefore, confidentiality is the most serious concern in cloud especially for data storage for instance customer data and credit card information [4], [29].As a

result, encryption is the fundamental solution to improve data, communications, processes confidentiality [30]. [31]studied the privacy and security issues in the cloud computing. However, the discussion of the security challenges was focused on confidentiality availability, integrity and accountability but they did not mention the limitations that caused the vulnerabilities such infrastructure design. This report aims to provide a step-by-step guideline for design and deployment as well as testing of a secure OpenStack cloud computing platform. discover and highlight the security issues and limitations that lead to security holes and vulnerabilities in cloud computing.

### 4.1.2 Integrity

According to [32] Data integrity is the assurance to eliminate third parties from auditing the data. Since the fundamental solution to ensure confidentiality cloud computing, hence, data integrity should be checked and maintained constantly to guarantee that data are accurate. Cloud providers should ensure that any modification to the data is detected. Integrity also covered that computer programs execution should be protected from malware, malicious user or insider, which may render an incorrect result or change the program execution[33]. Over the years, researchers have proposed different data integrity scheme, for instance, ithe authors in [34]highlights the importance of data integrity and presented a taxonomy of data integrity schemes that could be used in for cloud computing. However, Solutions proposed are not suitable for all types of data and environments.

### 4.1.3 Availability

Availability remains a challenge in cloud computing since cloud systems infrastructure are complex because it provides different services with different requirements. According to Khan et al. availability is calculated as the percentage of time an application and its services are available, given a specific time interval. Many services can be used to improve the availability of a service, such as, load balancing, and redundancy [35]

### 4.2 Insecure API

Service representation in cloud computing is given by client applications, supported by application program interface (API)[36] . In practical, API is a middleware that uses to integrate between underlying layers where developers use them for the application development and high- level applications layer [37]. Many researchers studied clouds and API security in particular, [38]summarized that APIs should provide flexible security interfaces to limit the existing security holes. One of the suggestion is that API should be provided with standards and security controls to enhance APIs with authentication and authorization using Open Authorization Protocol (OAuth). However, in 2015 Phogat and Sujatha studied OAuth authentication protocol mechanism in APIs, they conclude that OAuth security does not provides an acceptable security level. On the other hand, [39]highlighted that OAuth is an ideal candidate for Software as a Service, nevertheless different service provides use different version of OAuth and offer different authentication/authorization process which posing interesting challenges at the same time. Therefore, they proposed OAuthHub solution which used as a single trusted intermediary to control and manage data is shared and how authentication is done. In addition, the proposed solution required to isolate the APIs in the memory.  In fact, the cloud applications may have subjected to the same vulnerabilities as traditional Web applications. Nevertheless,[34] stated that the traditional security solutions are not sufficient for the cloud computing environment. This is probably driven from the fact that cloud computing vulnerabilities can lead to huge damage and propagate more widely. The distributed location of multiple users, resources and data make the cloud computing security a serious issue. Therefore, management, use and development of Web applications must consider web application vulnerabilities risk to protect the cloud. Insecure API can be disturbing for the users and the cloud. The APIs vulnerabilities includes insufficient authorization, input-data validation and weak credentials. However, after exhaustive research, it was founded that there is a lack of researches that consider API in open stack. Thus, this thesis focuses on implementing a penetration testing on API open stack core services.

## 4.3 Information security protection

Information security protection may be achieved through implementing technical, physical, management, and operational measures designed to protect the confidentiality, integrity and availability of information. [40]suggest that such control to consist of Detective, Preventative, Responsive and countermeasures such as Intrusion Detection Systems (IDS), Intrusion Prevention Systems (IPS) and firewalls. Effective information systems controls include multiple layers' approach, to mitigate the threats such an approach is very effective if one control is bypassed the other control can detect or prevent threat such as having Network based IDS and Host IDS.

Table 3 Comparison of firewalls advanced features

| FIREWALL | CHANGING DEFAULT POLICY TO ACCEPT/REJECT (BY ISSUING A SINGLE RULE) | ETHERNET MAC DESTINATION ADDRESS | ETHERNET MAC SOURCE ADDRESS | INBOUND FIREWALL (INGRESS) | CHANGE TTL? (TRANSPARENT TO TRACEROUTE) | CONFIGURE REJECTWITH ANSWER | AV (ANTIVIRUS) |
|---|---|---|---|---|---|---|---|
| PFSENSE | YES | YES | YES | YES | NO | YES | YES |
| UNTANGLE | YES | NO | NO | YES | NO | NO | YES |
| VYATTA | YES | NO | YES | NO | NO | NO | YES |
| IPFIRE | YES | YES | NO | YES | YES | NO | YES |

### 4.3.1 Firewall

Firewall as described by [41] is a network security device that monitors incoming and outgoing network traffic that can be a software program or a dedicated network appliance, which decides whether to allow or block specific traffic based on a defined set of security rules Since this project is focused on opens source solution, many open source firewalls have been looked at e.g. IPfire, Vyttta, Untangle and Pfsense. Based on the comparison of various features such as usability, advanced rule-set, rule-set Appliance-UTM filtering, base OS, pfsense is the preferred tool of choice because of its versatility.

Pfsense is an open source network firewall distribution, based on the FreeBSD operating system with a custom kernel which including third-party open source software packages for additional functionality pfsense.org [42]. Pfsense can implement the same performance as commercial firewalls, without any of the artificial limitations [42].

### 4.3.2 Intrusion Detection

An Intrusion Detection System (IDS) is a network security technology originally for detecting vulnerability exploits against a target application or computer. Intrusion Prevention Systems (IPS) extended IDS solutions by adding the ability to block threats in addition to detecting them [40], [43].
A vast amount of research has been carried by many researchers comparing the two best known open sources IDS/IPS snort and suricate. For instance,[40] looks at performance deference's between two IDS/IPS and suggest that snort is better. In addition [43] Compares the fail positives, their result suggests snort is better. For this project, snort will be used.

# 5 Step-by-Step OpenStack Deployment

As mentioned earlier, the aim of this report is to detail the steps required to perform a complete installation of OpenStack on multiple nodes. We split the installation process into two parts:
The first part is Deploying, securing BradStack multi-node OpenStack cloud infrastructure and developing OpenStack deployment toolkit.The second part is focused on conducting Penetration testing on OpenStack Services to validate the effectiveness of the security controls on the BradStack CCP.
This report describes how the BradStack OpenStack cloud infrastructure was designed, built and configured. The chapter also looks at the installation of the following security mechanisms; Pfsense firewall, snort (IPS/IDS) and Cisco switch hardening that include a step by step guide to using the proposed OpenStack installation toolkit.

## 5.1 Hardware Setup

Table 4 Hardware Setup and Specifications

| Component | Quantity | Description |
|---|---|---|
| Admin PXE node | 1 | Dell Optiplex 745<br>CPU: Intel Core 2 6400 @ 2.13GHz X 2 Cores<br>RAM: 2 GB<br>OS: 64 Bit<br>HDD: 160 GB<br>NIC: X 1 |
| Cloud Controller | 1 | Dell Precision T5400<br>CPU: Intel Xeon E5405 @ 2.00GHz X 8 Cores<br>RAM: 32 GB<br>OS: 64 Bit<br>HDD: 1TB<br>NIC: X 2 |
| Compute servers | 2 | Dell Precision T3400<br>CPU: Intel Core 2 Quad Q6600 @ 2.40GHz X 4 Cores<br>RAM: 4 GB<br>OS: 64 Bit<br>HDD: 500 GB<br>NIC: X 1 |
| Block Storage Server | 1 | Dell PowerEdge 1600SC<br>CPU: Intel Xeon @ 2.8GHz X 2 Cores<br>RAM: 4 GB<br>OS: 64 Bit<br>HDD: 150 GB |
| Network server | 1 | Dell Precision T3400<br>CPU: Intel Core 2 Quad Q6600 @ 2.40GHz X 4 Cores<br>RAM: 4 GB<br>OS: 64 Bit<br>HDD: 500 GB |
| Data switch | 1 | HP Procurved Networking 10Gbps |
| Private switch | 1 | ZyXEL Internet Security gateway |
| Cables |  | 7 x RJ 45 straight through copper cables |

## 5.2 Proposed BradStack Architectural Design

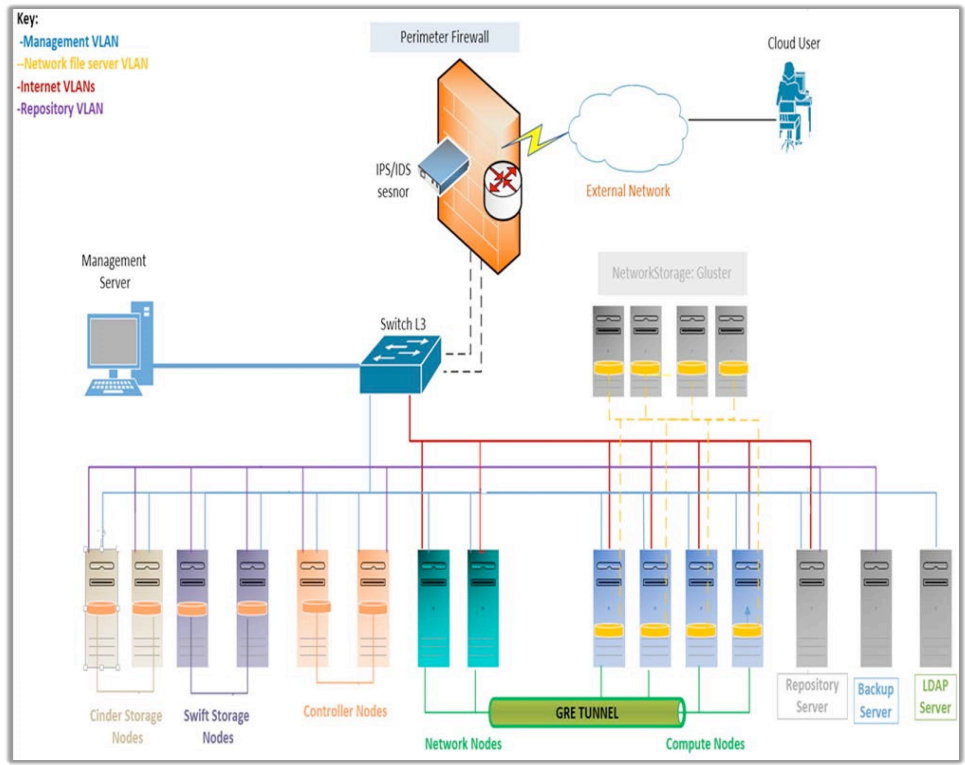

Figure 5.1 Proposed BradStack Design

Figure 5.1 depicts the proposed infrastructure topology (and architecture) of the redesigned, secured and reconfigured BradStack OpenStack CCP. The architecture comprises of the additional security layer using the PfSense firewall, snort IDS/IPS, tunnelling to segment networks as well as the failover strategy. Details of the architecture and how it's been implemented are provided in the next subsections.

## 5.3 Network Infrastructure Implementation and Configuration

The following sections looks at steps taken to install Pfsense, snort and Cisco switch hardening including their configurations.

## 5.4 Installing and Configuring Snort

This section evidences the installation configuration and updating of snort IPD/IDS. Snort has been discussed in chapter 2. Snort status can be found Figure 3-2 shows the status of the snort installed rules.

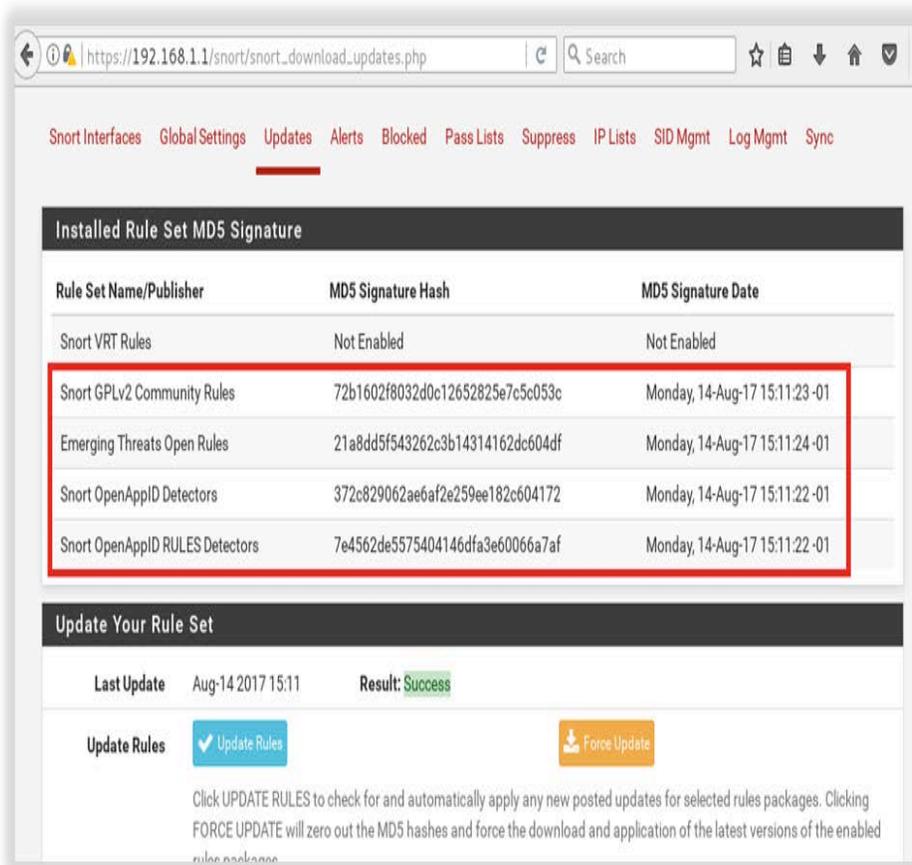

Figure 5.2 snort update status

## 5.5 Configuring Switch

This section looks at implementing Cisco's best practices for device hardening which include the following countermeasures; service password-encryption, BPDU guard, transport input SSH and switch port-security including mac-address sticky have been implemented and tested.

## 5.6 IP Addressing Table

Table 5 IP addressing table

| OPENSTACK SERVICE NAME | HOSTNAME | MANAGEMENT VLAN | REPOSITORY VLAN | INTERNET VLAN | NFS VLAN | TENNANT VLAN | TENNANT VLAN |
|---|---|---|---|---|---|---|---|
| **CONTROLLER** | Richmond1 | 192.168.1.100 | | 192.168.20.99 | | | |
| | Richmond2 | 192.168.1.101 | | 192.168.20.100 | | | |
| **NETWORK** | | 192.168.1.106 | | 192.168.20.110 | | 192.168.3.0/24 | 192.168.4.0/24 |
| | | 192.168.1.107 | | 192.168.20.111 | | | |
| **COMPUTE** | Horton1 | 192.168.1.110 | 192.168.2.101 | 192.168.20.101 | 192.168.10.101 | | |
| | Horton2 | 192.168.1.111 | 192.168.2.102 | 192.168.20.102 | 192.168.10.102 | | |
| | Horton3 | 192.168.1.112 | 192.168.2.103 | 192.168.20.103 | 192.168.10.103 | | |
| | Horton4 | 192.168.1.113 | 192.168.2.104 | 192.168.20.104 | 192.168.10.104 | | |
| **CINDER** | Phoenix1 | 192.168.1.120 | | 192.168.20.120 | | | |
| | Phoenix2 | 192.168.1.121 | | 192.168.20.121 | | | |
| **SWIFT** | Ashfield1 | 192.168.1.130 | | 192.168.20.130 | | | |
| | Ashfield2 | 192.168.1.131 | | 192.168.20.131 | | | |

# 6 BradStack Implementation and Configuration

This section gives a brief overview of OpenStack Alternative installer, OpenStack installation toolkit, BradStack OpenStack deployment process, BradStack OpenStack Cloud Infrastructure implementation and BradStack CCP cloud testing.

## 6.1 OpenStack Alternative Installer Comparison

**PackStack**: Packstack is an OpenStack installer that uses Puppet for deploying OpenStack services. Puppet modules deploy OpenStack components on multiple pre-installed servers over SSH automatically [22].

**DevStack** : DevStack is a sequence of expansible shell scripts used to rapidly install OpenStack environment depending on the latest versions of everything from git master. It is deployed as a development setting and as the basis for much of the OpenStack project's functional testing[22].

**ANVIL :** ANVIL is open source python shell scripts and utilities that can be used to deploy OpenStack. This tool is mainly for developers [22].

Table 6 depicts OpenStack alternative installer. It can be observed from the table below that DevStack is a developer installer and it's not stable. But ANVIL only supports older version of Centos, which is a big limitation and also, it's not stable and doesn't support multiple node installations. PackStack is also a developer installer which is not stable as the rest of the developer installers mentioned above. Juju and MaaS are commercial installation tools their biggest drawback is the difficulty level of the installation and also limits the number of nodes one can deploy for free. As stated by [44]This limitation of Juju and MaaS is similar when it comes to fuel Mirantis. However, our installer overcome all the limitations associated with the installation tools presented above with the additional advantages of modularization approaches which is capable of production environment deployment and easy to install.

Table 6 OpenStack Installer Comparison

| Features\installer | Devstack | ANVIL | PackStack | Juju and MaaS | Mirantis fuel | Proposed Installation toolkit |
|---|---|---|---|---|---|---|
| Deployment | Developer | Developer | Developer | Commercial | Commercial | Production environment |
| Supporting operating System | Ubuntu, fedora and centos RHEL7 | RHEL 6.2 Centos Oracle enterprise | Ubuntu Centos 7 | Ubuntu | Ubuntu | Centos |
| Support Multinode | yes | no | yes | yes | yes | yes |
| opensource | yes | yes | yes | yes | yes | yes |
| Stability of deployment Setup | no | no | Stable than devstack | yes | yes | yes |
| Difficulty level | no | no | normal | High | High | easy |

## 6.2 BradStack OpenStack Deployment Process

This deployment toolkit is based on OpenStack installation tutorials and guides. Even with the tutorial provided, installation is not that straightforward due to its complexity and the number of steps required to get fully functional Cloud infrastructure as discussed in the previous section. Figure 6.1 below illustrates the process for implementing BradStack OpenStack Octaca based on the development toolkit.

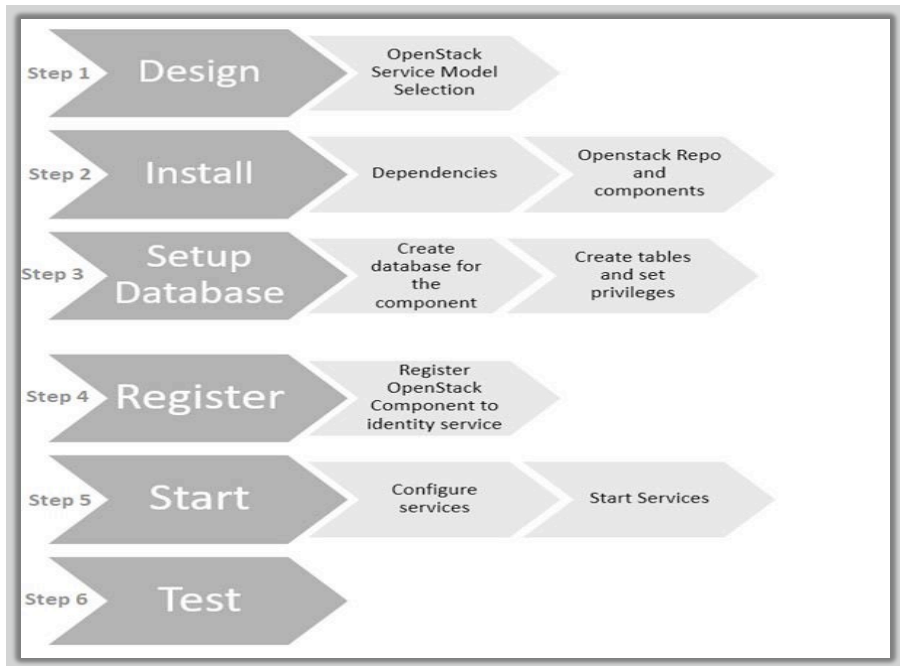

Figure 6.1 BradStack OpenStack deployment process

- **Design**

Involves selecting the operating systems, choosing OpenStack service model, installing and updating all nodes, configuring network interfaces and name resolution and conducting basic connectivity test.

- **Install**

At this step, all OpenStack Dependents and requirements are installed after adding Centos-Release-OpenStack-Ocata and epel7 repository. Each service has its own shell script to install all its requirements.

- **Database Setup**

This involves installing and creating database for all the selected services during the design stage in contrast to OpenStack guide where the database for each service is configured as the service is being installed.

- **Register**

This step involves the creation of domains, projects, users, and roles and all services selected during the design phase created including API endpoints. OpenStack uses three API endpoint variants for each service: admin, internal, and public endpoint. The admin API endpoint allows modifying users and tenants. More details on OpenStack API endpoints can be found in the next section.

- **Start**

This step involves installing and configuring service before starting them for instance Glance, Nova, Neutron and Cinder. This includes controller node and services in their relevant node such as Nova in compute node. This is dependent on the design stage what OpenStack Services are required and some core service that is required in any design.

- **Test**

At this step, all installed OpenStack services will be verified, followed by functionality testing which involves creating network and starting virtual machines.

## 6.3 BradStack OpenStack Implementation Process

In this section, we will cover the procedures to install OpenStack using the suggested toolkit and the process mentioned above. Each service has its own shell script in line with the installation process steps. Below is the flow and layout of the OpenStack installation toolkit.

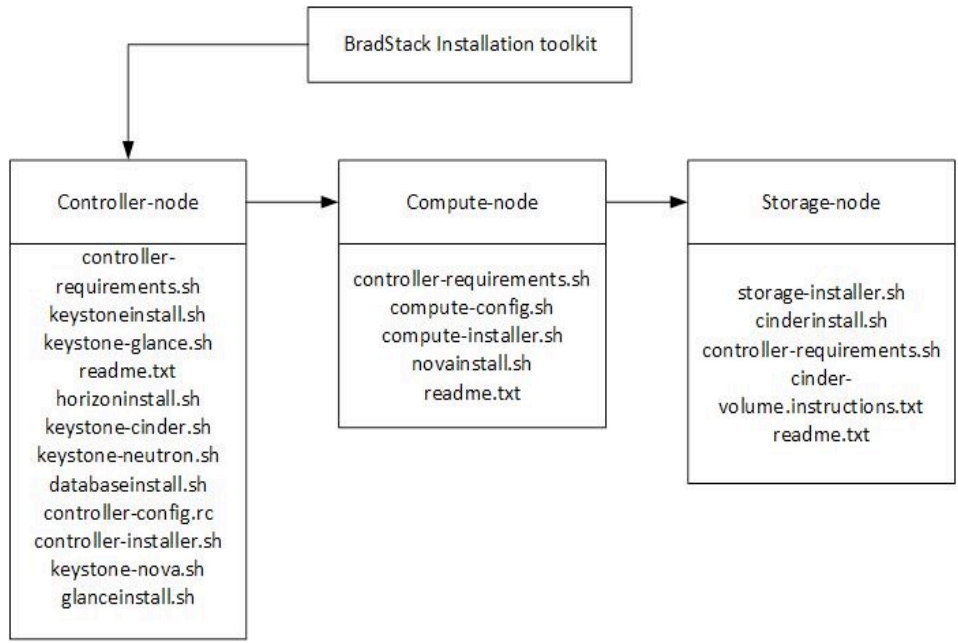

Figure 6.2 BradStack Installation toolkit layout

### 6.3.1 Step 1: Design (BradStack design)

The BradStack infrastructure Design is based on the BradStack requirement in terms of what services for example storage as a service detects the installing of swift. This service and their variations have been discussed in the previous chapter. Figure 6.3 below is the service design for BradStack. The controller box shows all the services installed on the controller node, Compute box show the services installed on the compute node, storage shows the services installed in the storage nodes and the block storage box shows all the service installed block storage nodes. Centos 7.3 has been chosen over Ubuntu as the base systems because Ubuntu tends to be less conservative than CentOS, which means that its repositories contain fresher packages, whereas CentOS packages tend to focus on security patches, stability and consistency.

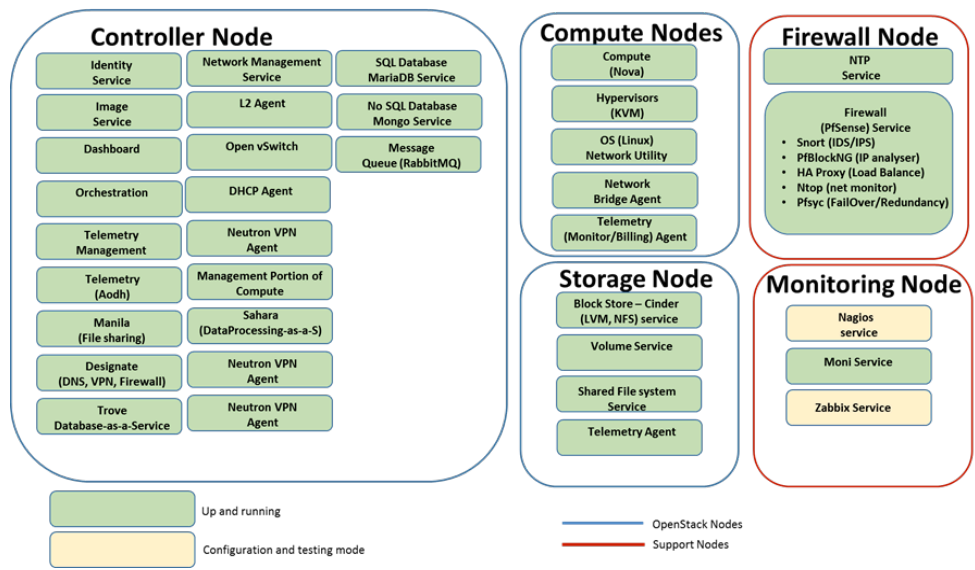

Figure 6.3 BradStack Service Design

### 6.3.2 Step 2: Install (installation requirements)

Adding centos-release-OpenStack, EPEL repository and install dependents on all nodes. All this done by a shell script, which is named requirements.sh. Each service has different requirements so thus, each module has one below are the three module and their requirement shell scripts.

**Controller-node:**
```
#the following must be run as root cd
/Desktop/deployment-kit/controller-node
./chmodcontroller  +x *.sh- requirements.sh
```

**Compute-node**
#the following must be run as root
```
cd /Desktop/deployment-kit/compute-node
chmod +x *.sh
./compute-requirements.sh
```

**Storage-node:**
#the following must be run as root
```
cd /Desktop/deployment-kit/storage-node
chmod +x *.sh
./storage-node-requirements.sh
```

### 6.3.3 Step 3: Setup database (services database creation)

This step only applies the controller nodes (Richmond). MariaDB-server-10.0, MariaDB-client was installed using the shell script named database.sh. This script creates all database for the services that have been selected at the design stage of BradStack. Below is the command in the database install scripts.

Install Database

```
yum -y erase mysql
yum -y install mariadb-galera-server mariadb-galera-common mariadb-galera galera
yum -y install OpenStack-utils
crudini --set /etc/my.cnf.d/server.cnf mysqld max_allowed_packet 256M           sed -i -r "s/^bind-address.*=.*0.0.0.0/bind-address=0.0.0.0\nmax_connections=$dbmaxcons/" /etc/my.cnf.d/galera.cnf
systemctl enable mariadb.service
systemctl start mariadb.service
/usr/bin/mysqladmin -u $mysqldbadm password $mysqldbpassword > /dev/null 2>&1
/usr/bin/mysqladmin -u $mysqldbadm -h $dbbackendhost password $mysqldbpassword > /dev/null 2>&1
```

After installing the database software then followed by services databases are installed starting with Keystone, Nova, Glance, Neutron and cinder. Only keystone script snippet will be shown below due to document constants. The full script will be available made available online for reference.

```
echo "CREATE DATABASE $keystonedbname default character 
set utf8;"|$mysqlcommand         echo "GRANT ALL ON 
$keystonedbname.* TO '$keystonedbuser'@'%' 
IDENTIFIED BY 
'$keystonedbpass';"|$mysqlcommand
 echo "GRANT ALL ON $keystonedbname.* TO 
'$keystonedbuser'@'localhost' 
IDENTIFIED BY '$keystonedbpass';"|$mysqlcommand
echo "GRANT ALL ON $keystonedbname.* TO 
'$keystonedbuser'@'$keystonehost' 
IDENTIFIED BY 
'$keystonedbpass';"|$mysq
lcommand         for 
extrahost in 
$extrakeystonehosts
        do
            echo "GRANT ALL ON $keystonedbname.* TO 
'$keystonedbuser'@'$extrahost' 
IDENTIFIED BY '$keystonedbpass';"|$mysqlcommand
        done
        echo "FLUSH PRIVILEGES;"|$mysqlcommand
```

### 6.3.4 Step:4 Register (service registration and identities)

This step will install keystone, bootstrap keystone, create services, add a member and all the service from design step will have a service created and endpoint. Using script keystone.sh. This toolkit has been designed in a modular approach, which means for each service has its own keystone install script, for instance, keystone-glance, which would create glance service and the API endpoints.

```
Installing keystone
cd 
/Desktop/deployment-
kit/controller-node
chmod ./core +-
xservices *.sh 
/install-keystone.sh

yum -y install OpenStack-keystone OpenStack-utils 
OpenStack-selinux python-psycopg2
yum -y install mod_wsgi memcached 
python-memcached httpd yum -y 
install python-OpenStackclient
# We also start/enable memcached service
cat ./libs/memcached/memcached 
> /etc/sysconfig/memcached
systemctl enable memcached
systemctl stop memcached
systemctl start memcached
```

```
# Using pyhton based "ini" configuration tools, we
begin Keystone configuration crudini --set
/etc/keystone/keystone.conf DEFAULT compute_port
8774 crudini --set /etc/keystone/keystone.conf
DEFAULT debug False crudini --set
/etc/keystone/keystone.conf DEFAULT log_file
/var/log/keystone/keystone.log
crudini --set /etc/keystone/keystone.conf DEFAULT
use_syslog False crudini --set
/etc/keystone/keystone.conf memcache servers
$keystonehost:11211
```

After installing keystone service users are created, roles and endpoints for Nova, Glance, Neutron and cinder on snippet for neutron will be shown below

```
.source $keystone_admin_rc_file echo ""

echo "Creating
NEUTRON
Identities"
echo ""
#"Neutron User:"
OpenStack user create --domain $keystonedomain --password
$neutronpass --email
$neutronemail $neutronuser
#"Neutron Role:"
OpenStack role add --project $keystoneservicestenant --user
$neutronuser
$keystoneadminuser
#"
Neutron Service:" OpenStack service create \
        --name $neutronsvce \
        --description "OpenStack Networking" \
        network
#"Neutron Endpoints:"
OpenStack endpoint create -
-region $endpointsregion \
network public
http://$neutronhost:9696
OpenStack endpoint create -
-region $endpointsregion \
network internal
http://$neutronhost:9696
OpenStack endpoint create -
-region $endpointsregion \
network admin
http://$neutronhost:9696
```

### 6.3.5 Step: 5 Start (starting the OpenStack services)

In this step, all services will be started then in the next step verified in the next step.

```
cd /Desktop/deployment-kit/ ./OpenStack-
services-restart.sh
```

### 6.3.6 Step: 6 Test (service verification)

This section will involve verifying that all services are running, creating a network, flavors, uploading an image to glance and finally launching the virtual machine. Only Nova and cinder service will be displayed due to document constraints. The following will involve querying service to view their status. Figure 6.4 below shows the status of services.

**Nova**

Figure 6.4 Nova service list

**Cinder**

Figure 3-6 below show the status of cinder services.

Figure 6.5 Cinder Service List

# 7 BradStack Cloud Testing

In this section, the deployed BradStack cloud infrastructure will be extensively tested. First, a private network is created, upload image then launching an instance. The image below shows the network topology from BradStack functionality test

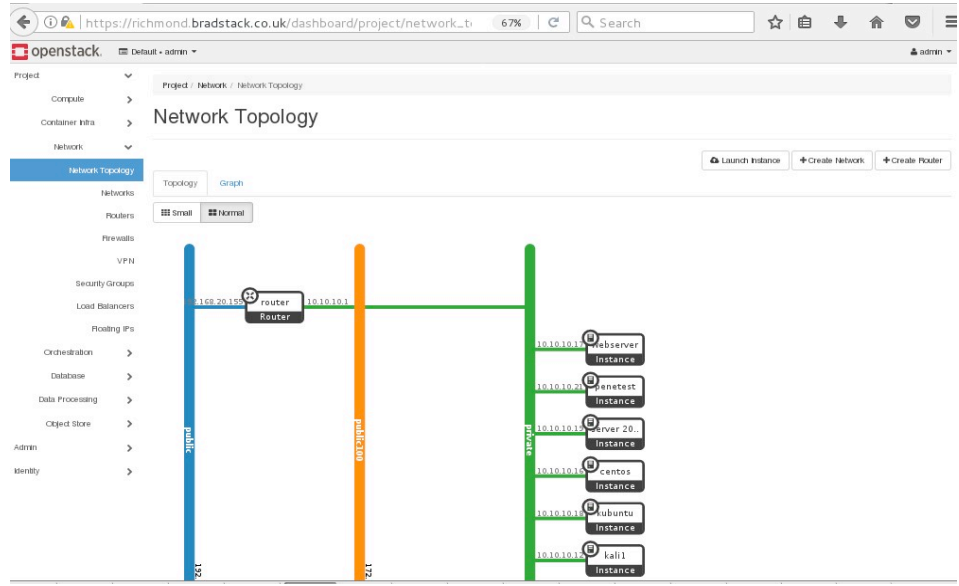

Figure 7.0 Network Topology with running instances

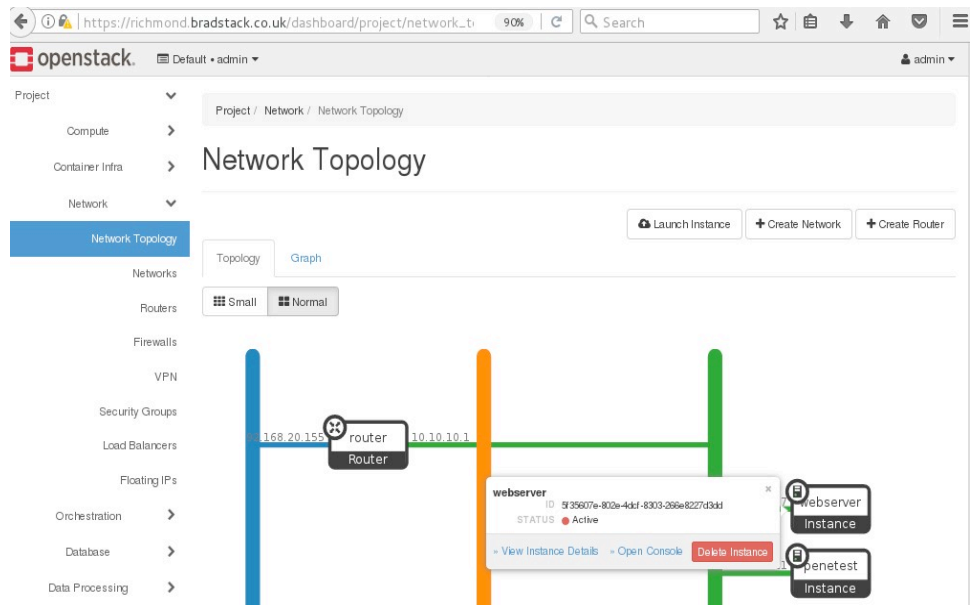

Figure 7.1 Network Topology with webserver and pentest running instance

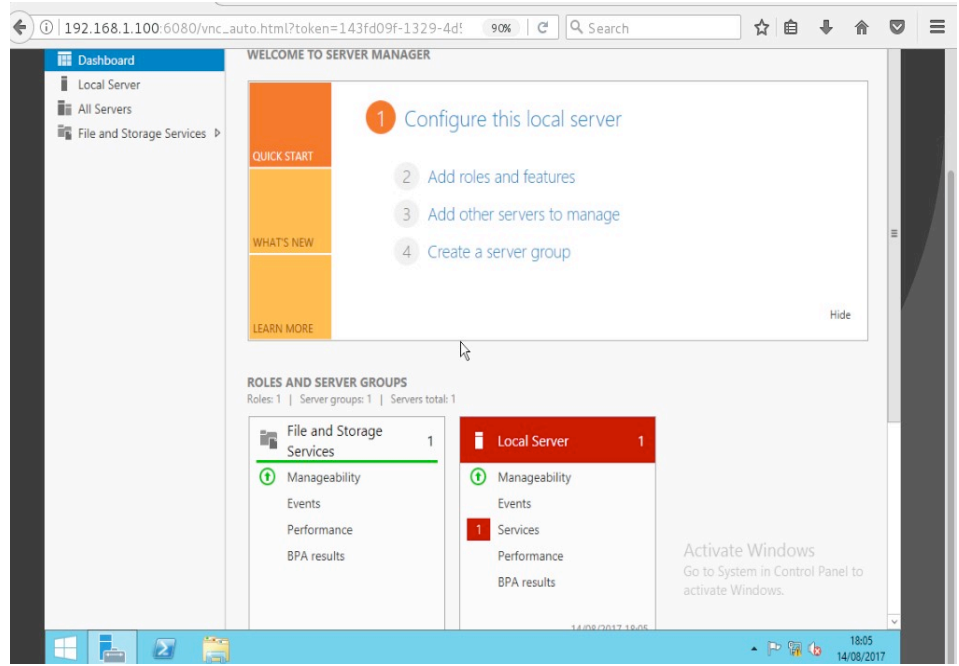

Figure 7.2 Network Topology with windows Server running instance

## 7.1 BradStack OpenStack Installer

The following is the database-install script

## Database-install.sh

```bash
#!/bin/bash
#

PATH=$PATH:/bin:/sbin:/usr/bin:/usr/sbin:/usr/local/bin:/usr/local/sbin

#
# First, we source our config file.
#
if [-f
./conf/controller
-config.rc ] then
source
./conf/controller
-config.rc
else
echo "Can't access controller-config.rc configuration file. Aborting !"    echo ""    exit 0 fi
```

```bash
echo "Installing Local MariaDB Software"
rm /root/.my.cnf
    yum -y erase mysql
yum -y install mariadb-galera-server mariadb-galera-common mariadbgalera galera
yum -y install OpenStack-utils
crudini --set /etc/my.cnf.d/server.cnf mysqld max_allowed_packet 256M
sed -i -r "s/^bind-address.*=.*0.0.0.0/bind-address=0.0.0.0\nmax_connections=$dbmaxcons/" /etc/my.cnf.d/galera.cnf
systemctl enable mariadb.service
systemctl start mariadb.service
/usr/bin/mysqladmin -u $mysqldbadm password $mysqldbpassword > /dev/null 2>&1
/usr/bin/mysqladmin -u $mysqldbadm -h $dbbackendhost password $mysqldbpassword > /dev/null 2>&1
sleep 5
echo "[client]" > /root/.my.cnf
echo "user=$mysqldbadm" >> /root/.my.cnf
echo "password=$mysqldbpassword" >> /root/.my.cnf
echo "GRANT ALL PRIVILEGES ON *.* TO '$mysqldbadm'@'%' IDENTIFIED BY '$mysqldbpassword' WITH GRANT OPTION;"|mysql
echo "FLUSH PRIVILEGES;"|mysql
iptables -A INPUT -p tcp -m multiport --dports $mysqldbport -j ACCEPT
service iptables save
echo "MariaDB Installed"
# The following two variables are used later in the database creation section
#
mysqlcommand="mysql --port=$mysqldbport --password=$mysqldbpassword --user=$mysqldbadm --host=$dbbackendhost"
echo "[client]" > /root/.my.cnf
echo "user=$mysqldbadm" >> /root/.my.cnf
echo "password=$mysqldbpassword" >> /root/.my.cnf
```

```
echo "Keystone"
echo "CREATE DATABASE $keystonedbname default
character set utf8;"|$mysqlcommand         echo
"GRANT    ALL   ON   $keystonedbname.*    TO
'$keystonedbuser'@'%'
```

## 7.2 Creating private network

Figure 7.3 below shows the creation of private software-defined network (SDN).

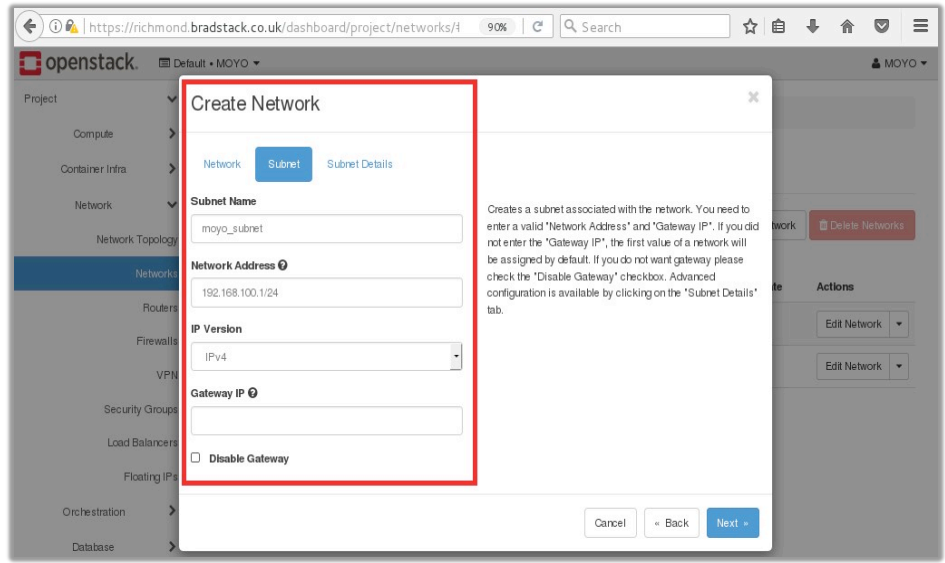

Figure 7.3 Private Network Creation

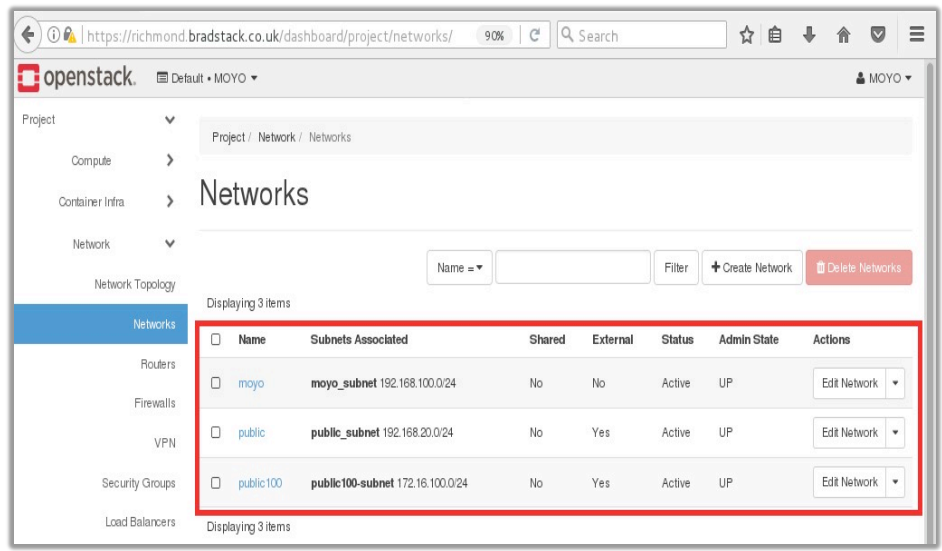

Figure 7.4 Moyo Network Overview

Figure 7.4 below shows an overview of the entire network with Moyo account which includes the public networks that are used for floating allocation

## 7.3 Building an instance

The following section demonstrates the brief steps to launching of an instance from web application Horizon. The following information is required; image, flavour which specifies the how resources the instance needs, for example, random access memory (RAM), virtual processing unit (VPU), Operating System image and a network.

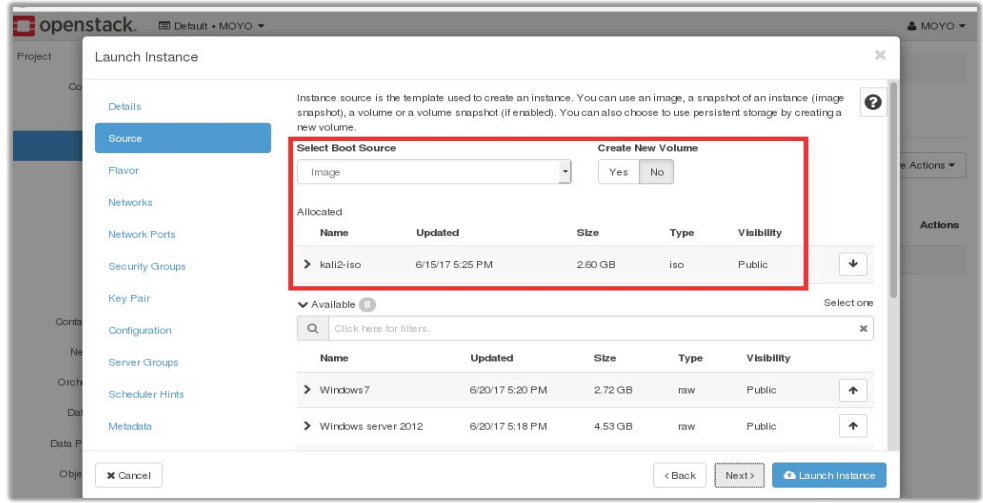

Figure 7.5 selecting Kali Linux image as the boot source

Figure 7.5 shows the selection of the image to be booted and figure 7.6 shows the status of the launched instance.

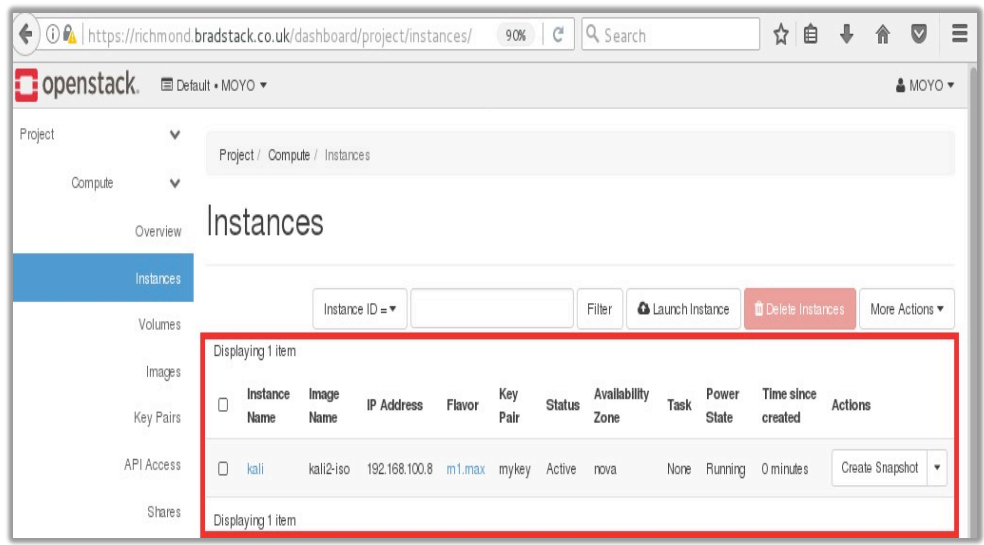

Figure 7.6 Status of Kali instance

Figure 7.7 below shows the Kali Linux instance full desktop experience of the test instance kali. The building and deploying of the instance took a few minutes which shows how fast the infrastructure is in terms of performance.

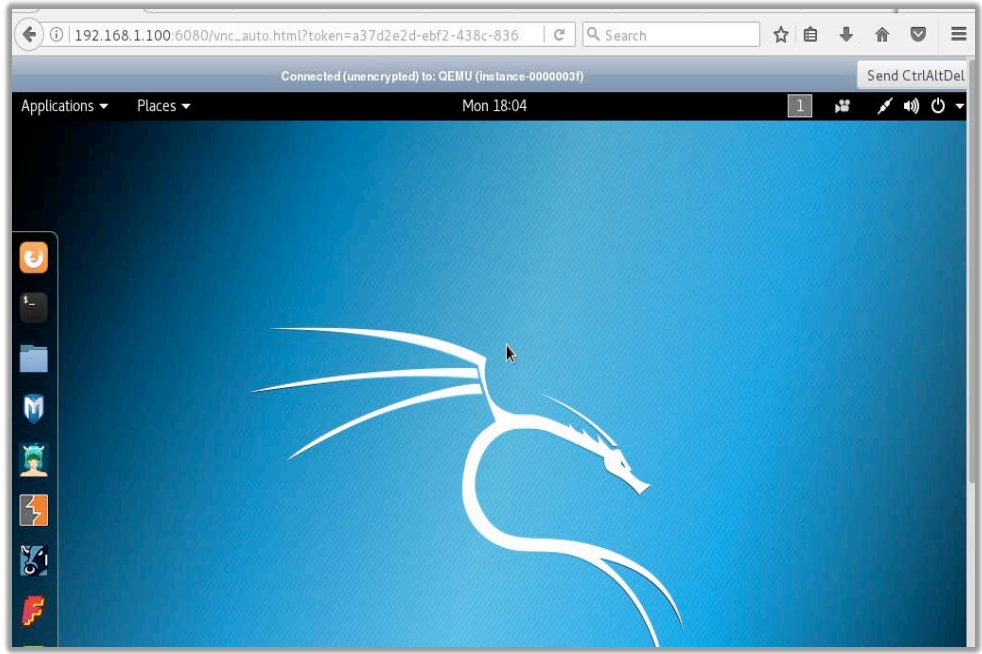

Figure 7.7 Kali Linux full desktop

# 6 Conclusions

Cloud computing's growing popularity especially using infrastructure as a service has inspired many academic institutions to transform their current infrastructure into a private or hybrid cloud. However, even though openstack was not the first to propose open source cloud computing frameworks, but it works with popular enterprise and open source technologies making it ideal for heterogeneous infrastructure. In this report, three issues have been successfully addressed which include: securing BradStack's network infrastructure, configuring and deploying BradStack using a developed toolkit as well as testing the entire cloud key functionalities. We divided our proposed strategy into two parts: The first part involves redesign, redeploying of BradStack CCP using the developed OpenStack installation toolkit, and the second part looks at testing all the cintslled cloud services.This report serves as a practical guideline to academic instituons intending to build a private cloud and focusing on security and practical infrastructure related issues. It also serves as a reference for institutions looking at the possibilities of implementing a secured cloud solution.

In the future we will focus on securing BradStack multi-node OpenStack cloud infrastructure by incorporating network firewalls, intrution detection and prevention sytems, web application firewall, network segregation through the use of virtual LANs and implementing Cisco's best practise on switch hardening